\documentclass[fleqn,usenatbib]{mnras}
\usepackage[T1]{fontenc}
\DeclareRobustCommand{\VAN}[3]{#2}
\let\VANthebibliography\thebibliography
\def\thebibliography{\DeclareRobustCommand{\VAN}[3]{##3}\VANthebibliography}
\usepackage{graphicx}
\usepackage{amsmath}
\usepackage{mathtools}
\usepackage{amssymb}
\usepackage{ulem}
\usepackage{comment}
\usepackage{color,hyperref}
\usepackage{fontawesome}
\usepackage{threeparttable}
\usepackage{newtxtext,newtxmath,times,verbatim}
\usepackage{booktabs}

\newcommand{\pro}{\textsc{Prospector}}

\title[Dwarf AGN from SED fitting]{The properties of AGN in dwarf galaxies identified via SED fitting}

\author[B. Bichang'a et al.]{B. Bichang'a,$^{1}$\thanks{E-mail: b.o.bichanga@herts.ac.uk} S. Kaviraj,$^{1}$ I. Lazar,$^{1}$ R. A. Jackson,$^{2}$ S. Das,$^{1}$ D. J. B. Smith,$^{1}$ A. E. Watkins,$^{1}$ G. Martin$^{3}$\\
$^{1}$Centre for Astrophysics Research, University of Hertfordshire, Hatfield AL10 9AB, UK\\
$^{2}$Department of Physics and Astronomy, University of Victoria, BC V8P 5C2, Canada\\
$^{3}$School of Physics and Astronomy, University of Nottingham, University Park, Nottingham NG7 2RD, UK\\
}

\begin{document}
\label{firstpage}
\pagerange{\pageref{firstpage}--\pageref{lastpage}}
\maketitle

\begin{abstract} 
Given their dominance of the galaxy number density, dwarf galaxies are central to our understanding of galaxy formation. While the incidence of AGN and their impact on galaxy evolution has been extensively studied in massive galaxies, much less is known about the role of AGN in the evolution of dwarfs. We search for radiatively-efficient AGN in the nearby ($0.1<z<0.3$) dwarf (10$^{8}$ M$_{\odot}$ < $M_{\rm{\star}}$ < 10$^{10}$ M$_{\odot}$) population, using SED fitting (via \textsc{Prospector}) applied to deep ultraviolet to mid-infrared photometry of 508 dwarf galaxies. Around a third (32 $\pm$ 2 per cent) of our dwarfs show signs of AGN activity. We compare the properties of our dwarf AGN to control samples, constructed from non-AGN, which have the same distributions of redshift and stellar mass as their AGN counterparts. KS tests between the AGN and control distributions indicates that the AGN do not show differences in their distances to nodes, filaments and nearby massive galaxies from their control counterparts. 
This indicates that AGN triggering in the dwarf regime is not strongly correlated with local environment. The fraction of AGN hosts with early-type morphology and those that are interacting are also indistinguishable from the controls within the uncertainties, suggesting that interactions do not play a significant role in inducing AGN activity in our sample. Finally, the star formation activity in dwarf AGN is only slightly lower than that in their control counterparts, suggesting that the presence of radiatively-efficient AGN does not lead to significant, prompt quenching of star formation in these systems. 
\end{abstract}

%The AGN and controls do not show strong differences in their distances to nodes, filaments and nearby massive galaxies, {\color{red}with KS tests between the AGN and control distributions returning p-values much larger than 5 per cent}. 


%%%%%%%%%%%%%%%%%%%%%%%%%%%%%%%%%%%%%%%%%%%%%%%%%%

%%%%%%%%%%%%%%%%% BODY OF PAPER %%%%%%%%%%%%%%%%%%

\section{Introduction}
\label{sec:introduction}

The steep rise in the galaxy stellar mass function in the dwarf ($M_{\rm{\star}}$ < 10$^{10}$ M$_{\odot}$) regime \citep[e.g.][]{Wright2017,Martin2019,Driver2022} indicates that dwarf galaxies dominate the galaxy number density at all redshifts and in all environments. This makes them vital, both for our empirical understanding of galaxy evolution and for constraining theoretical models. Our current picture of how galaxies evolve is largely based on massive galaxies, since these objects are bright enough to be detectable, across a large range in redshift, in past wide-area surveys \citep[e.g.][]{Jackson2021a}.

The properties of dwarf galaxies have been studied in detail in our local neighbourhood, out to distances of $\sim$50 Mpc \citep[e.g.][]{Mateo1998,Tolstoy2009}. However, typical dwarfs become undetectable at cosmological distances, in past surveys like the Sloan Digital Sky Survey \citep[SDSS;][]{York2000,Alam+2015}, which offer large footprints but are relatively shallow \citep[e.g.][]{Jackson2021a,Lazar2024}. In the specific case of the SDSS, it is worth noting that, while the limiting magnitude of its standard depth imaging is $r\sim$22.7, many scientific analyses, including past dwarf studies using the SDSS, use the spectroscopic main galaxy sample (MGS). The MGS, however, has a much shallower magnitude limit of $r\sim$17.77 \citep{Strauss2002}. The dwarfs that exist in datasets like the SDSS MGS tend to have high star formation rates (SFRs), which boost their luminosities above the detection thresholds of shallow surveys but also bias them towards systems that are blue, star-forming and potentially dominated by late-type morphologies \citep[e.g.][]{Lazar2024}. 

The advent of new surveys that are both deep and wide, like the recent Hyper Suprime-Cam Subaru Strategic Program \citep[HSC-SSP;][]{Aihara2018} and, in the near future, Rubin Observatory's LSST \citep{Ivezic2019}, offer optical photometry with unprecedented depth over large areas of the sky. Such datasets can be used to identify dwarfs down to $M_{\rm{\star}}$ $\sim$ 10$^{8}$ M$_{\odot}$ out to at least $z\sim0.3$ \citep{Jackson2021a}. In conjunction with similarly deep surveys at ancillary wavelengths, such datasets are likely to transform our understanding of the dwarf regime outside the very local Universe. It is worth noting that the shallow potential wells of dwarf galaxies are more sensitive to key processes that affect galaxy evolution (e.g. ram pressure stripping, tidal interactions, and baryonic feedback from stars and AGN) than massive galaxies \citep{Martin2019}. Statistical studies of dwarfs outside the local neighbourhood are, therefore, likely to lead to significant gains in our understanding of the physics of galaxy evolution. 

A prominent topic in the recent literature is the presence of black holes (BHs) and the role of AGN in galaxy evolution. The massive galaxy regime is well studied in this context, with both observations and theory suggesting that not only do all massive systems likely host black holes \citep{Richstone1998}, but that star formation activity, and therefore the growth of the host galaxy, is heavily influenced by AGN feedback \citep[e.g.][]{Fabian2012,Beckmann2017,Kaviraj2017,Davies2023}. However, less is currently known about the presence of BHs in dwarfs and their potential role in driving galaxy evolution in this regime \citep[e.g.][]{Silk2017,Volonteri2021}. While there is accumulating observational evidence of AGN in dwarf galaxies \citep[e.g.][]{Greene2007,Reines2013,Marleau2017,Nyland2017,Satyapal2018,Mezcua2018,Mezcua2019,Kaviraj2019,Baldassare2020,Davis2022}, questions persist regarding their frequency, the processes that might trigger the BHs, correlations with local environment and whether feedback from the AGN can have a similar impact in dwarfs as it does in the massive-galaxy regime. It is worth noting that, in the dwarf regime, BHs tend not to grow in current cosmological simulations. This is because the BHs either wander for significant fractions of their lifetimes in regions of low gas density, away from the barycentres of the galaxies, or because supernova feedback displaces material around the BH and and prevents it from feeding \citep[e.g.][]{Dubois2021}. However, it remains unclear whether this behaviour truly mimics that of real dwarf galaxies or may be a result of the characteristics of such simulations (e.g. their relatively low mass and spatial resolutions). Empirical constraints on the incidence of AGN in dwarfs and the properties of their host galaxies are, therefore, valuable. 

While AGN emission contributes to all parts of the spectral energy distribution (SED), the samples of AGN recovered from probing specific wavelength ranges differ, with unique challenges experienced in the different methods that are employed. For example, AGN selection based on X-ray luminosity \citep[e.g.][]{Latimer2019}, may not be ideal for identifying AGN in the dwarf regime, due to the long exposure times required to detect low-luminosity AGN \citep[e.g.][]{Reines2020}. %The similar colour ratios of starburst galaxies and AGN, may limit the effectiveness of mid-infrared selection techniques \citep[e.g.][]{Mezcua2018}. 
AGN selection in dwarfs via optical variability \citep[e.g.][]{Baldassare2020,Martinezpalomera2020}, ideally requires optical surveys that are both deep and wide, to ensure that the dwarfs being studied do not suffer from the biases in shallow surveys described above. Employing emission-line diagnostics \citep[e.g.][]{Baldwin1981,Kewley2001,Cidfernandes2010,Cidfernandes2011} to detect AGN in dwarfs in shallow surveys may similarly miss some AGN. This is because, if dwarfs in such surveys are biased towards star-forming systems then the impact of the AGN on the emission lines may be dominated by the star formation activity. 

Several studies in the literature have used radio excess to detect AGN \citep[e.g.][Drake et al. in preparation]{Lofthouse2018,Gurkan2018}. However, an implementation of this technique on data from the LOFAR Two-metre Sky Survey \citep{Shimwell2019} indicates that, given the lower radio luminosities of dwarf AGN, only young radio AGN are likely to be found in dwarf galaxies \citep[e.g.][]{Davis2022}, even in the deepest high-resolution radio surveys that are currently available.  

In this context, SED fitting \citep[e.g.][]{Leja2018}, using deep multi-wavelength broadband photometry is a useful avenue for identifying AGN in dwarf galaxies \citep[e.g.][]{das2024lofar,Delvecchio2014,Delvecchio2017,Thorne2022,Best2023}. Using the full SED simultaneously leverages the impact of the AGN over a range of wavelengths. Furthermore, since high signal-to-noise ratio observations of fainter objects like dwarfs are easier to achieve via broadband photometry, more complete samples of dwarfs can be probed for the presence of AGN, at least in the nearby Universe. The purpose of this paper is to use deep broadband multi-wavelength photometry to search for dwarfs which exhibit signs of AGN at low redshift ($z<0.3$), probe the role of interactions and environment in AGN triggering and study whether there is evidence for significant prompt quenching of star formation in these systems. 

This paper is organized as follows. In Section \ref{sec:data} we describe the broadband photometric data that is used in the SED fitting, the calculation of environmental parameters, the selection of our sample of dwarf galaxies and the morphological classification of these systems using \textit{Hubble Space Telescope} (HST) images. In Section \ref{sec:prospector}, we describe the calculation of physical parameters using the SED-fitting code \textsc{Prospector} \citep{Johnson2019,Leja_2019,Johnson_2021}. In Section \ref{sec:results}, we use the outputs of our SED fitting to explore the incidence of AGN in our dwarf galaxies, study the role of interactions and local environment in AGN triggering and compare the star formation activity in AGN and non-AGN to explore the possibility of AGN-driven quenching. We summarise our findings in Section \ref{sec:summary}. 

%.............................................................

\section{Data}
\label{sec:data}

\subsection{The COSMOS2020 catalogue}
\label{sub:cosmos2020} 

We employ photometric data from the Classic version of the COSMOS2020 catalogue \citep{Weaver+2022}, which provides deep multi-wavelength ultra-violet (UV) to mid-infrared (MIR) photometric data in the COSMOS field. Notable improvements in this catalogue compared to its predecessor, the COSMOS2015 catalogue \citep{Laigle_2016}, is the incorporation of ultra-deep optical ($griz$) data from the Hyper Suprime-Cam (HSC) Subaru Strategic Program \citep{Aihara2018} for object detection, improved Spitzer/IRAC data processing using \textsc{IRACLEAN} and improved astrometric precision from aligning the imaging data with the Gaia DR1 \citep{GaiaCollaboration_2016}. The optical HSC photometry has a point source depth of $\sim$28 mag, significantly deeper than past surveys such as the SDSS (in which the optical images are $\sim$5 mag shallower than those from the HSC in COSMOS). We refer readers to \cite{Weaver+2022} for further details of the processing of the photometric data. 

We use total magnitudes from 15 broadband filters that span the full UV to MIR wavelength range provided by the catalogue from the following surveys: $FUV$ and $NUV$ data from GALEX \citep{Zamojski_2007}, $u$/$u^*$ from MegaCam/CFHT \citep{Sawicki_2019}, $g$, $r$, $i$, $z$, $y$ from Subaru/Hyper Suprime-Cam \citep{Aihara_2019}, $Y$, $J$, $H$ and $K_s$ from VIRCAM/VISTA \citep{McCracken_2012} and mid-infrared data from IRAC/Spitzer Channels 1, 2, 3 and 4 \citep{Ashby_2013,Steinhardt_2014,Ashby_2015,Ashby_2018}. The optical and infrared photometry is extracted using the \textsc{SExtractor} and \textsc{IRACLEAN} codes respectively. Since we are specifically interested in dwarfs, which have smaller angular sizes, the high resolution of the IRAC MIR imaging \citep[2 arcsec;][]{Fazio_2004} is an advantage.   

In the $u$-band footprint, the catalogue offers photometry from two virtually identical filters, $u$ and $u^*$. The $u$ filter has a slightly shorter central wavelength and has replaced $u^*$ in recent CFHT surveys, such as the Large Area U-band Deep Survey \citep[CLAUDS;][]{Sawicki_2019}. While the two filters span a similar wavelength range, the $u$ filter does not cover the full COSMOS field. $u^*$ data is deeper in the central region than the $u$ band which, however, is more uniform across the field. For our work, we use the deeper of the two bands when both are available. The filterset available spans emission from starlight through to the hot dust that is likely to be associated with the AGN \citep[e.g.][]{Hickox2018}. The IRAC wavebands are able to provide significant discrimination between the MIR emission produced by stellar populations from that originating from the AGN \citep[e.g.][]{Assef2010}. 

\begin{figure*}
\center
\includegraphics[width=2\columnwidth]{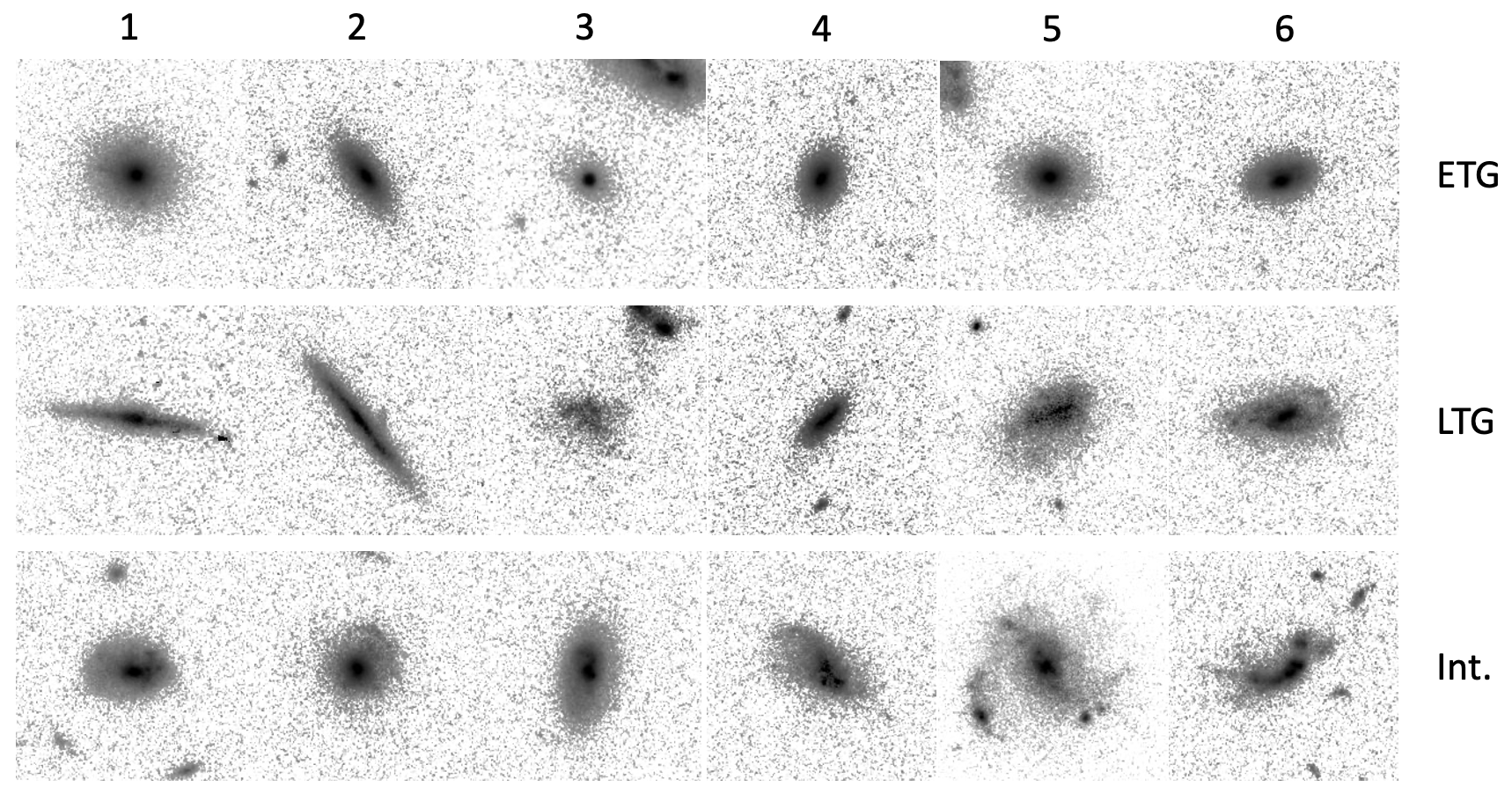}
\caption{Example HST/F814W images of dwarf galaxies in our sample, including ETGs (top row), LTGs (middle row) and interacting systems (bottom row). In the bottom row, the images in the first three columns show examples of interacting ETGs, while the last three columns show examples of interacting LTGs. Each image has a size of 5 arcseconds on a side. At the median redshift of our dwarf sample ($z\sim0.21$), this corresponds to a physical size of around 18 kpc.}
\label{fig:images}
\end{figure*}

%.............................................................

\subsection{Environmental parameters}
\label{sub:density_maps}

Since we are interested in probing the role of environment in AGN triggering, we estimate environmental parameters using the DIScrete PERsistent Structure Extractor \citep[\textsc{DisPerSE};][]{Sousbie_2011}. Following \citet{Laigle_2018} and \citet{Lazar_2023}, we use \textsc{DisPerSE} to construct topological maps of the large-scale structure within the COSMOS footprint, in our redshift range of interest\footnote{\citet{Libeskind+2018} provide a comprehensive comparison of filament finding algorithms including \textsc{DisPerSE}.}.  

\textsc{DisPerSE} computes a discrete Morse complex, using the Delaunay tessellation field
estimator \citep[DTFE;][]{Schaap+2000,Cautun+2011} to create a density map. It first uses galaxy positions to compute Delaunay tessellations, i.e. the triangulation of discrete points such that no points from the defining distribution are within any of the resulting triangles. It then computes the density ($\delta$) of the resulting cells, where $\delta$ = (area of cell)$^{-1}$, and identifies the locations of stationary points (i.e. maxima, minima and saddle points). Finally, from this computed density field, segments are used to connect the nodes with the saddle points. The set of segments represents the filamentary structure of the cosmic web.  

\textsc{DisPerSE} utilises the concept of persistence to ensure the robustness of its density maps. It defines a persistence parameter ($N$) %, expressed as a multiple of standard deviations ($\sigma$), 
which serves as a threshold for the critical pairs that contribute to the density map and is important for constructing a robust filament structure devoid of noisy structures. The algorithm only retains critical pairs which have Poisson probabilities above $N\sigma$ from the mean. Following \citet{Laigle_2018} and \citet{Lazar_2023}, we set $N$ equal to 2. We use massive ($M_{\rm{\star}}$ > 10$^{10}$ M$_{\odot}$) galaxies to build our density maps, as they dominate the local gravitational potential, with the thicknesses of the slices ($\delta z \sim 0.03$) being driven by the median redshift uncertainties of the massive galaxies.   

%.............................................................

\subsection{Sample Selection}
\label{sub:selection}

As noted in the introduction, given the depth of the photometry in COSMOS2020, complete samples of dwarfs can be constructed down to $M_{\rm{\star}}$ $\sim$ 10$^8$ M$_{\odot}$, out to at least $z\sim0.3$ \citep[e.g.][]{Jackson2021a}. For our study, we leverage the COSMOS2020 catalogue to pre-select our galaxies as follows. Tests indicate that, for the same set of UV -- MIR photometric data points, the stellar masses outputted by \textsc{Prospector} are slightly larger than those produced by \textsc{LePhare} (which underpins COSMOS2020) by around 0.3 dex (this trend has also been noted by \citet{Pacifici+2023}). Our pre-selected sample consists of COSMOS2020 galaxies at $0.1<z<0.3$ in which the 16\textsuperscript{th} percentile (i.e. the -1$\sigma$) value of the probability density function (PDF) of the stellar mass is greater than 10$^{7.5}$ M$_{\odot}$ and the 84\textsuperscript{th} percentile (i.e. the +1$\sigma$) value is less than 10$^{9.5}$ M$_{\odot}$. 

There are {\color{black}$\sim$6500} such COSMOS2020 objects which are both classified as galaxies in this catalogue (\texttt{lp\_type} = 0) and as `extended' (i.e. galaxies) in the $g$, $r$, $i$ and $z$ bands by the HSC pipeline and which lie outside masked regions such as bright star masks and image edges. Of these, {\color{black}1146} galaxies have the full complement of UV to MIR photometry. These galaxies form the initial sample of objects on which we perform SED fitting using \pro, as described in Section \ref{sec:prospector} below. For each galaxy, we use the photometric redshift from the COSMOS2020 catalogue and the median stellar mass as a first guess in the SED fitting process. As described in Section \ref{sec:results} below, when identifying AGN and studying their properties, we refine the sample further, by applying additional cuts on redshift accuracy and the quality of the SED fits from \pro.  

%.............................................................

\subsection{Morphological classification using HST/F814W images}
\label{sub:visual_classification}

We use visual inspection of HST F814W ($I$-band) images of the COSMOS field \citep{Koekemoer2007,Massey2010} to morphologically classify our dwarf galaxy sample. These images have a 5$\sigma$ point-source depth of 27.8 mag and an angular resolution of $\sim$0.05 arcseconds. The visual inspection is performed by one expert classifier (SK). We classify our dwarfs into two broad morphological classes: early-type galaxies (ETGs) and late-type galaxies (LTGs). We also flag galaxies that show evidence of an ongoing or recent interaction e.g. tidal features, internal asymmetries or tidal bridges due to an ongoing merger with another galaxy. Figure \ref{fig:images} presents example images of our dwarfs, including ETGs (top row), LTGs (middle row) and interacting systems (bottom row). Images in the first three columns in the bottom row show examples of interacting ETGs, while the last three columns show examples of interacting LTGs.  

%.............................................................

\section{SED fitting using Prospector}
\label{sec:prospector}

We use \textsc{Prospector}, a \textsc{Python}-based SED-fitting code to fit the photometry of our dwarf sample. \textsc{Prospector} is a Bayesian forward-modelling code \citep[see Table 1 in][for a list of popular codes in the current literature]{Pacifici+2023} that can be used to infer galaxy properties, using either photometric or spectro-photometric data. \textsc{Prospector} is able to fit either parametric or non-parametric star formation histories (SFHs) to galaxy data and simultaneously constrain the AGN contribution to the SED. The code uses the Flexible Stellar Population Synthesis package \citep[\textsc{FSPS};][]{Conroy+2009} to generate SED models on the fly using \textsc{Python-FSPS} \citep{dan_foreman_mackey_2014}. 

We sample the posterior parameter distribution using \textsc{Dynesty} \citep{Speagle_2020}, which utilises the dynamic nested sampling algorithms of \citet{Higson_2019}. We adopt the non-parametric \textsc{continuity-sfh} models within \textsc{Prospector} as a basis for our analysis. Compared to their parametric counterparts, non-parametric models can typically explore more complex SFHs, which can include features like starbursts and episodes of quenching and rejuvenation \citep[e.g.][]{Simha_2014,Diemer_2017,Leja_2019,Johnson_2021}, which are challenging to explore using parametric models \citep[e.g.][]{Suess_2022}. 

The parameters in the \textsc{Prospector} model that we use for our analysis are briefly described below and summarised in Table \ref{tab:priors_table}. 

 \begin{table*}
	\centering
	\begin{tabular}{lrrrr} % four columns, alignment for each
		\hline
		Parameter  & Prior& Description\\
        \hline
        \hline
        $\log (M_\star$/M$_{\odot}$)        & Uniform (min = $5$, max = $13$) & Stellar mass\\
        $\log r_{\rm i}$                              & Student's t-distribution & Ratio of SFRs in adjacent bins\\
        $\log (Z_\star / Z_\odot)$                              & Uniform (min =$-2.0$, max = $0.2$) & Stellar metallicity\\
        \hline
        $\tau_{\rm 5500,diffuse}$                             & Clipped normal (mean = $0.3$, std. dev. = $1.0$, min = $0$, max = $4$) & Optical depth of diffuse dust\\
        $\tau_{\rm young}/\tau_{\rm diffuse}$                  & Clipped normal (mean = $1$, std. dev. = $0.3$, min = $0$, max = $1.5$) & Ratio of optical depth of birth clouds to optical depth of diffuse dust\\
        $\Gamma_{\rm dust}$                                    & Clipped normal (mean = $0$, std. dev. = $0.5$, min = $-1$, max = $0.4$) & Delta of the slope of attenuation curve compared to Calzetti law\\
        \hline
        $U_{\rm min}$                                    & Uniform (min = $0.1$, max= $25$) & Minimum intensity of light affecting the dust\\
        $Q_{\rm PAH}$                                            & Uniform (min = $0.5$ , max= $7$) & Fraction of dust mass in PAHs\\
        $\log \gamma_{\rm dust}$                             & Uniform (min = -3, max = 0.15) & Fraction of dust mass exposed to $U_{\rm min}$\\
        \hline
        $\log f_{\rm AGN}$                                 & Uniform (min = -5, max = 3) & AGN luminosity as a fraction of the galaxy's bolometric luminosity\\
        $ \log \tau_{\rm AGN}$         & Uniform (min = 5, max = 150) & Optical depth of the AGN torus\\
        \hline
        $\log U_{\rm gas}$                                 & Uniform (min = -4, max = 1) & Gas phase ionization parameter\\
        \hline
	\end{tabular}
 \caption{SED-fitting parameters, and their associated priors, in our \textsc{Prospector} model.}
\label{tab:priors_table}
\end{table*}

\begin{itemize}

    \item \textbf{SFH:} We implement \textsc{continuity-sfh} by adopting six age bins divided logarithmically, except the first two age bins which are fixed at 0 -- 30 Myr, 30 -- 100 Myr and the final age bin which is fixed to cover 15 percent of the age of the Universe. We utilise the Student's t-distribution for the prior (the log of the ratio of the SFRs in adjacent bins) with $\mu$ = 2 (where $\mu$ is the number of degrees of freedom) and a scale-factor ($\sigma$) of 0.3.

    \item \textbf{Stellar mass:} We adopt a uniform prior for the log of the stellar mass in the range 5 < log ($M_{\star}$/\rm M$_{\odot}$) < 13. 

    \item \textbf{Stellar metallicity:} 
    A uniform prior is adopted for the log of the stellar metallicity in the range -2.0 < log ($Z_{\star}/\rm Z_{\odot}$) < 0.2 following \cite{Leja_2017}. This range brackets the stellar metallicities observed in nearby dwarf galaxies \citep[e.g.][]{Gallazzi_2005,Panter2008}.

    \item \textbf{Dust attenuation:} Following \cite{Johnson_2021}, we implement the two-component \citet{Charlot_2000} dust model that treats the effect of dust from birth clouds (which impacts young stars) separately from that due to the diffuse dust (which impacts light from both young and old stars). Following \citet{Kriek+2013}, the strength of the UV bump is tied to the power-law index of the curve (because steeper attenuation laws tend to have stronger bumps). As described in Table \ref{tab:priors_table}, the three parameters used to describe dust attenuation are the optical depth of the diffuse dust ($\tau_{\rm 5500,diffuse}$), the ratio of the optical depth of the birth clouds to that due to the diffuse dust ($\tau_{\rm young}$/$\tau_{\rm diffuse}$), and a delta modifier ($\Gamma_{\rm dust}$) to the slope of the Calzetti dust law \citep{Calzetti+2000}.
       
    \item \textbf{Dust emission:} The dust templates adopted \citep{Draine_2007} include infrared emission from small and large dust grains and polycyclic aromatic hydrocarbons (PAHs). The model includes three free parameters: $U_{\rm min}$, $\gamma_{\rm dust}$ and $Q_{\rm PAH}$. $\gamma_{\rm dust}$ describes the fraction of dust mass affected by the minimum intensity of starlight required to heat up the diffuse dust and interstellar medium in the galaxy ($U_{\rm min}$). $Q_{\rm PAH}$ describes the fraction of dust mass contributed by PAHs, which are responsible for strong emission in the mid-infrared wavelengths. A uniform prior is adopted for all three free parameters within the ranges shown in Table \ref{tab:priors_table}.

    \item \textbf{AGN templates:} The AGN templates adopted by \textsc{Prospector} are based on the \textsc{CLUMPY} models \citep{Nenkova2008a,Nenkova2008b} and include two free parameters: $f_{\rm{AGN}}$ and $\tau_{\rm{AGN}}$. $f_{\rm{AGN}}$ describes the ratio of the AGN luminosity to  the galaxy's bolometric luminosity, while the latter describes the optical depth of the AGN torus. We adopt a uniform prior in the log of these parameters, within the ranges shown in Table \ref{tab:priors_table}. 

    \item \textbf{Nebular emission:} Nebular emission in \textsc{Prospector} uses the \textsc{FSPS} implementation of the \textsc{CLOUDY} photoionization code \citep{Ferland2013}, where the ionising sources are the stellar populations produced by FSPS, as described by \citet{Byler2017}. The model is parametrized by the gas phase metallicity and the ionization parameter, $U_{\rm gas}$ \citep{Leja_2017}. Following the default implementation within \pro, we tie the gas-phase metallicity to the stellar metallicity and allow the log of the ionization parameter to vary between -4 and 1. 
    
\end{itemize}

We consider the acceptability of our fits following \citet[][]{Smith+2012}. A fit is considered to be acceptable if its best-fit $\chi^2$ is below the 99 percent confidence threshold for the given number of photometric bands. We fit each galaxy with and without the AGN templates. We use the ratio of the $\chi^2$ of the best-fit model with and without the AGN templates to estimate the significance of including an AGN in the fit. %We use the ratio of the $\chi^2$, {\color{black}computed using \pro's prediction of the SED for the highest probability sample}, with and without the AGN to estimate the significance of including an AGN in the fit. 
As described in Section \ref{sec:agn_fraction} below, we then use that information, together with the value of $f_{\rm AGN}$, to identify dwarfs which are likely to show signs of AGN activity. 

%.............................................................

\section{AGN in nearby dwarf galaxies}
\label{sec:results}

In this section we use the outputs from our SED fitting to probe the incidence of AGN in nearby dwarfs, study the role of interactions and environment in AGN triggering and compare the star formation activity in dwarf AGN to their non-AGN counterparts. For the analysis below, we restrict our initial sample of {\color{black}$1146$} galaxies to those which have photometric redshift errors less than 20 per cent and have acceptable fits as described above. This produces a final sample of {\color{black}508} dwarf galaxies that underpins our analysis below. Of these, {\color{black}476} galaxies (i.e. around 94 per cent) lie within the HST footprint and our morphological analysis in Section \ref{sub:morphology} is therefore restricted to these systems. It is worth noting that, by construction, the AGN that we study here are only those that are radiatively-efficient and leave a signature in the UV to MIR wavelengths. The stellar mass and SFR values used in our analysis below are the median values of these parameters, calculated using \textsc{Prospector}, for each galaxy. 

%.............................................................

\subsection{The incidence of AGN in dwarf galaxies}
\label{sec:agn_fraction}
 
\begin{figure}
    \centering
    \includegraphics[width=\columnwidth]{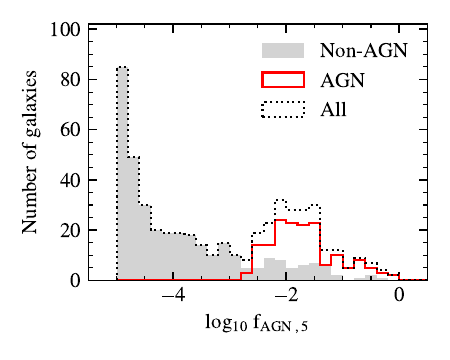}
    \caption{Distribution of $f_{\rm{AGN,5}}$ in our final sample of 508 dwarf galaxies, which exhibits a bimodality around $\log f_{\rm{AGN,5}}\sim -2.8$. The red histogram denotes galaxies that we define to be AGN. These are galaxies which have $\log f_{\rm{AGN,5}} > -2.8$ and in which including the AGN in the fitting improves the quality of the fit (i.e. the $\chi^2$ value of the best-fit model with an AGN template included is smaller than its counterpart without an AGN). The grey histogram corresponds to galaxies that are classified as non-AGN. These are systems that either have $\log f_{\rm{AGN,5}} < -2.8$ or where $\log f_{\rm{AGN,5}} > -2.8$ but including the AGN does not result in a better fit (see the text in Section \ref{sec:agn_fraction} for more details).}
    \label{fig:agn_bimodal}
\end{figure}

While we use the median values of stellar mass and SFR, we employ the 5\textsuperscript{th} percentile values of $f_{\rm{AGN}}$, denoted by $f_{\rm{AGN,5}}$ hereafter, to identify dwarfs which show signs of AGN activity in our sample. This ensures that 95 per cent of the $f_{\rm{AGN}}$ distribution in each galaxy is above the chosen threshold that defines an AGN. This is more conservative than using, for example, the 16\textsuperscript{th} percentile value, as in other recent studies \citep[e.g.][]{das2024lofar,Best2023}. Figure \ref{fig:agn_bimodal} presents the values of $f_{\rm{AGN,5}}$ in our final sample of dwarfs. The bimodality of the distribution (around $\log f_{\rm{AGN,5}}\sim -2.8$) suggests the presence of a population of AGN in these dwarf galaxies that could be demarcated by this value. We use this distribution to define AGN as galaxies where (1) $\log f_{\rm{AGN,5}} > -2.8$ and (2) including the AGN in the fitting improves the quality of the fit (i.e. the $\chi^2$ value of the best-fit model with AGN templates included is smaller than its counterpart without an AGN). The red histogram denotes the galaxy population that we define to be AGN in our sample. The grey histogram denotes non-AGN i.e. galaxies that either have $\log f_{\rm{AGN,5}} < -2.8$ or have $\log f_{\rm{AGN,5}} > -2.8$ but where including the AGN templates does not result in a better fit.  

\begin{figure}
     \centering
         \centering
         %\includegraphics[width=\columnwidth]{images/results/sample_heatmap_fagn16.pdf}
         % \caption{f$_{\rm{AGN\ (5)}}$}
         \centering
         \includegraphics[width=\columnwidth]{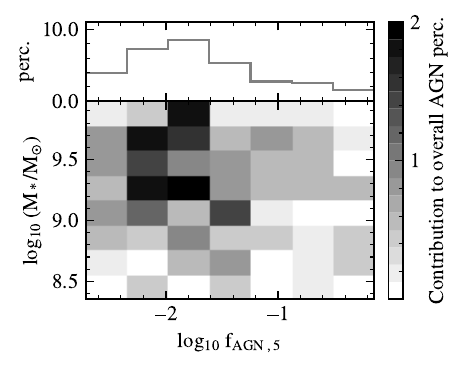}
         % \caption{f$_{\rm{AGN(16)}}$}
         \label{fig:heatmap_16}
        \caption{The contribution of dwarf AGN in different parts of the \textsc{Prospector}-derived stellar mass vs $f_{\rm{AGN,5}}$ parameter space to the overall AGN percentage in our dwarf sample. The histogram shows how this contribution varies as a function of $f_{\rm{AGN,5}}$ (regardless of the value of stellar mass). Summing the numbers in either the heatmap or the histogram yields the overall AGN percentage of $\sim$32 per cent.} 
        \label{fig:heatmapfagn_vs_mass}
\end{figure}

\begin{table*}
\begin{center}
\begin{tabular}{ l | r | r}
%\multicolumn{3}{c}{\textbf{Morphological fractions}}\\
\hline
& AGN                & Control\\
\hline
\hline

\begin{comment}
None $ 3.16 \pm 0.42 $
None $ 2.56 \pm 0.24 $
None $ 0.76 \pm 0.13 $
None $ 0.84 \pm 0.11 $
None $ 0.53 \pm 0.09 $
None $ 0.48 \pm 0.08 $
\end{comment}

Median projected distance to nearest node (Mpc) & $ 3.16 \pm 0.42 $ & $ 2.56 \pm 0.24 $ \\
Median projected distance to nearest filament (Mpc) & $ 0.76 \pm 0.13 $ & $ 0.84 \pm 0.11 $\\
Median projected distance to nearest massive galaxy (Mpc) & $ 0.53 \pm 0.09 $ & $ 0.48 \pm 0.08 $ \\
\hline
Fraction of galaxies which are ETGs         & $0.30 \pm 0.04$ & $0.33 \pm 0.04$ \\
Fraction of galaxies that are interacting & $0.14 \pm 0.03$ & $0.16 \pm 0.03$\\
\hline
Median log (sSFR / yr$^{-1}$) & $-10.37 \pm 0.05$ & $-10.22 \pm 0.04$ \\
\end{tabular}

\caption{Comparisons of various quantities between dwarf AGN and control samples constructed from the non-AGN dwarfs which have the same distributions of redshift and stellar mass as the AGN. Note that, while the analysis in rows 1--3 is restricted to the redshift range $0.2<z<0.3$ (because, as noted in Section \ref{sub:agn_env}, there are not enough massive galaxies to create reliable density maps at $z<0.2$), the analysis in rows 4-6 uses the full redshift range of our study ($0.1<z<0.3$). Rows 1--3 compare the distances to nodes, filaments and massive galaxies of the AGN and their control counterparts. The number of AGN on which this environmental analysis is based is {\color{black}63}. Rows 4 and 5 compare the fractions of AGN and control galaxies that are classified as ETGs and interacting respectively. The number of AGN which have HST images and underpin the analysis in these rows is {\color{black}155}. The uncertainties in rows 1--5 are calculated following \citet{Cameron2011}. Row 6 compares the median specific SFR (sSFR) of dwarf AGN to their control counterparts. The number of AGN that underpins this analysis is {\color{black}162}. The uncertainties in this row are calculated via bootstrapping.}
\label{tab:comparison}
\end{center}
\end{table*}

It is worth noting that the typical spatial scales of accretion disks are around $\sim$10$^{-2}$ pc \citep[e.g.][]{Hawkins2007}. This is around 5 orders of magnitude smaller on a linear scale (and therefore around 15 orders of magnitude smaller in terms of volume) than the host dwarf galaxies, which have spatial scales of around a kpc \citep[e.g.][]{Watkins2023}. Values of $\log f_{\rm{AGN,5}}$ values greater than -2.8 (i.e. $f_{\rm{AGN,5}}$ values greater than $\sim$0.16 per cent) may, therefore, represent relatively bright radiatively-efficient AGN in our dwarf population. Given the selection criteria above, around a third {\color{black}(162 out of 508 or 32 $\pm$ 2 per cent)} of the dwarfs in our sample show signs of AGN activity. 

In Figure \ref{fig:heatmapfagn_vs_mass}, we show the contribution of dwarf AGN in different parts of the $f_{\rm{AGN,5}}$ vs stellar mass space to the overall percentage of galaxies classified as AGN in our dwarf sample. The histogram shows how this contribution varies as a function of $f_{\rm{AGN,5}}$ (regardless of the value of stellar mass). Summing the numbers in the heatmap or the histogram yields the overall AGN percentage ({\color{black}around 32 per cent}) within our dwarf sample. Recall that the galaxy population studied here is likely to be complete down to $M_{\rm{\star}}$ $\sim$ 10$^{8}$ M$_{\odot}$, out to $z\sim0.3$ \citep{Jackson2021a}.%There is some evidence that more massive dwarfs contribute more to the overall AGN population, suggesting that the incidence of AGN may increase with stellar mass. %
%{\color{black}It is worth noting here that requiring a detection in IRAC Channels 2, 3 and 4, which trace hot dust, may bias our sample towards galaxies which host AGN in the first place. As a result, the AGN percentage derived here may be overestimated.} {\color{red}Let's think about whether this is actually true. Given the comments on the first draft this statement may not be well-motivated. I have removed it from the abstract for the time being.}

While past studies have estimated BH occupation and active fractions in dwarfs via different methods, our estimate for the AGN fraction in dwarfs is consistent with the findings of the recent dwarf literature. Recent work indicates that the BH occupation fraction in dwarfs is likely to be high \citep[estimates typically vary between 40 and 100 per cent within the stellar mass range probed in this study, see e.g.][]{Ricarte2018,Bellovary2019,Nguyen2019}. Inevitably, the active fractions are lower. 

For example, \citet{Mezcua2024} use spatially-resolved emission-line diagnostics from integral-field spectroscopy to estimate an AGN fraction of 20 per cent in SDSS dwarfs. This is consistent with the fractions found using mid-infrared photometry \citep[10--30 per cent, e.g.][]{Kaviraj2019}. It is interesting to note that the AGN fractions derived using a spaxel-by-spaxel analysis, as performed by \citet{Mezcua2024}, are much larger than those derived using single fibre diagnostics in the SDSS ($\sim$ 1 per cent). This is because, as noted in the introduction, the star formation activity across the galaxy is likely to swamp the emission-line signal due to the AGN in the highly star-forming dwarfs that are detectable in the SDSS. A spatially-resolved analysis, on the other hand, is able to isolate the region that may host an AGN, thus enabling a more sensitive probe of its existence within the galaxy. Interestingly, \citet{Dickey2019} use a sample of dwarfs with a median redshift of $z\sim0.024$ to show that, in the very nearby Universe -- where relatively unbiased dwarf samples can be constructed even using the SDSS -- around $\sim$80 per cent of quiescent dwarf galaxies in low-density environments show central AGN-like emission-line ratios. 

\citet{Pacucci_2021} estimate that  5 -- 22 per cent of the BHs in the dwarf population are likely to be active. This rises to around 30 per cent if the host metallicities are lower ($Z<\rm Z_{\odot}$), which is indeed the case for dwarf galaxies \citep[e.g.][]{Gallazzi_2005,Panter2008}. \citet{Davis2022} combine deep radio and optical data, from LOFAR and HSC respectively, with theoretical models to conclude that AGN triggering in dwarfs is likely to be stochastic and a common phenomenon. This appears consistent with the observation that dwarfs may lie on an extrapolation of the $M$ -- $\sigma$ relation seen in massive galaxies \citep[e.g.][]{Schutte2019,Davis2020}. Since BH feedback predicts a universal M $\propto$ $\sigma^4$ relation \citep[e.g.][]{Silk1998,King2021}, this suggests that frequent AGN activity could be important in the dwarf regime, consistent with the incidence of AGN being derived by recent work, including this study. Taken together, these studies suggest a growing consensus that a significant minority of dwarfs in the nearby Universe show signs of AGN activity. 
 
%.............................................................

\subsection{The role of environment in triggering dwarf AGN}
\label{sub:agn_env}

We proceed by comparing the environments of our dwarf AGN to their non-AGN counterparts. For this analysis, we consider galaxies in the redshift range $0.2<z<0.3$ because the relatively small number of massive galaxies at $z<0.2$ means that reliable density maps cannot be created for these redshifts. To perform this comparison, we construct a control sample of dwarfs in which $\log f_{\rm{AGN,5}} < -2.8$ and which has the same distributions of stellar mass and redshift as their dwarf AGN counterparts\footnote{We construct the control sample as follows. For each AGN we identify all non-AGN with $\log f_{\rm{AGN,5}} < -2.8$ within stellar mass and redshift tolerances of 0.05 dex and 0.04 respectively. We then select, at random, one of these galaxies to be the control counterpart of the AGN in question. No control galaxy is assigned to more than one AGN. The results in this section do not change if different tolerances are used.}. 

\begin{figure}
	\includegraphics[width=\columnwidth]{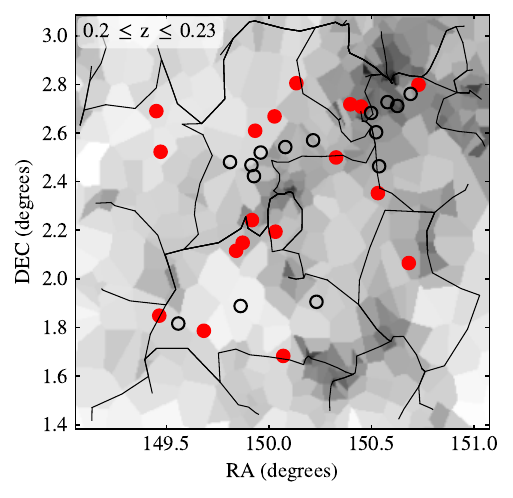}
    \caption{An example density map, generated using \textsc{DisPerSE}, as described in Section \ref{sub:density_maps} The redshift limits of the map are shown in the top left-hand corner. Filaments are shown using solid lines, with a random subset of AGN (filled red circles) and non-AGN (open black circles) shown overplotted.}
    \label{fig:density_maps}
\end{figure}

\begin{figure}
	\includegraphics[width=\columnwidth]{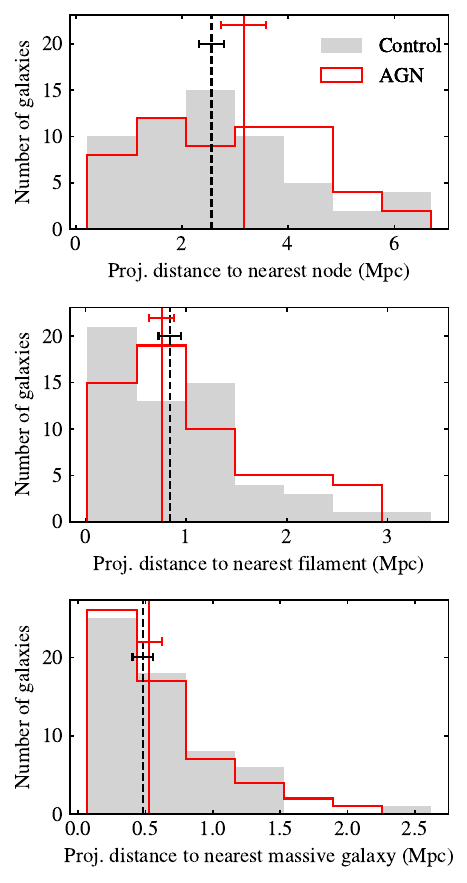}
    \caption{Projected distances to nearest nodes (top), filaments (middle) and massive galaxies (bottom) for dwarf AGN (red) and the control sample of non-AGN (grey).}    
    \label{fig:distance}
\end{figure}

In Figure \ref{fig:density_maps}, we show an example density map, created using \textsc{DisPerSE} as described in Section \ref{sub:density_maps}. The filaments are shown using solid lines and random examples of dwarf AGN and control galaxies that reside within the redshift slice are indicated using filled red and open black circles respectively. In Figure \ref{fig:distance}, we present the distributions of the projected distances to the nearest filaments, nodes and massive galaxies for the AGN and the control sample. Median values and their associated uncertainties, calculated via bootstrapping, are shown using the vertical lines and also described in rows 1, 2 and 3 of Table \ref{tab:comparison}. 

Applying a two-sample Kolmogorov-Smirnov (KS) test to the AGN and control distributions produces $p$-values of 0.25, 0.45 and 0.99 for the distances to nodes, filaments and massive galaxies respectively. Based on both the median values and their uncertainties and the results of the KS tests, we do not find evidence of a difference in these distances between the AGN and their control counterparts. This suggests that the environment in which they reside are unlikely to play a significant role in AGN triggering.

While various studies have probed the role of environment using different methods, our conclusions appear consistent with the results of other recent work which has explored the dwarf regime. For example, \citet{Kristensen2020}, who define local environment using the distance to the tenth nearest neighbour, find that environment is not an important factor in triggering AGN activity in dwarfs. This is similar to the results of \citet{Sabater2015}, whose galaxy sample includes objects at the upper mass end of the dwarf regime. They find that the effect of either the local density or of galaxy interactions is minor in the prevalence of AGN activity. 

%.............................................................

{\color{black}\subsection{Morphology and the role of interactions in AGN triggering}
\label{sub:morphology}

In this section we consider the morphological properties of our dwarf AGN compared to their non-AGN counterparts. Recall from Section \ref{sub:morphology} that morphological information is available for 476 galaxies out of the final sample of 508 which lie within the HST footprint. In a similar vein to Section \ref{sub:morphology} above, we construct a control sample which is matched in stellar mass and redshift to the AGN (but this time without any restriction in redshift).

In rows 4 and 5 of Table \ref{tab:comparison}, we summarise the fraction of AGN and control galaxies that are classified as ETGs and interacting respectively. The ETG and interacting fractions in the AGN and control samples are indistinguishable, given the uncertainties. The similar interacting fractions, combined with the fact that interactions reinforce dispersion-dominated ETG-like structures over time \citep[e.g.][]{Martin2018_sph}, suggests that this process does not play a significant role in triggering the AGN in this study. This result appears aligned with the findings of \citet{Kaviraj2019}, who have shown that the fraction of dwarf AGN in ongoing mergers does not appear to show an excess compared to that in their control counterparts. It is also consistent with the fact that AGN appear not to preferentially occupy regions of higher density (from Section \ref{sub:agn_env} above) where the incidence of interactions is likely to be higher. Finally, it is worth noting that the lack of influence of either local environment or the presence of interactions appears consistent with the conclusions of \citet{Davis2022} that AGN triggering in the dwarf regime may be a largely stochastic process.

%.............................................................

\subsection{Star formation activity in dwarf AGN}
\label{sub:star_formation}

We conclude our study by exploring whether there are differences in the star formation activity of dwarfs which contain AGN and those which do not. In particular, it is interesting to explore whether, given their relatively shallow potential wells, dwarf galaxies may be more susceptible to prompt quenching of star formation than their massive counterparts. In row 6 of Table \ref{tab:comparison}, we compare the median values of the specific SFRs (sSFRs), and their associated uncertainties, in dwarf AGN with those of the control sample defined in Section \ref{sub:morphology}. Given the median values and their uncertainties, the star formation activity in the AGN is marginally lower compared to those in their control counterparts. However, we do not find evidence of a significant, prompt quenching of star formation in dwarf galaxies which host AGN activity.    

It is worth discussing this result in the context of AGN feedback suppressing star formation in dwarf galaxies, the potential for which has been demonstrated from both theoretical \citep{Barai2019,Koudmani2022,ArjonaGalvez2024} and empirical \citep[e.g.][]{Penny2018} points of view. Recent work has shown that the presence of AGN feedback does not necessarily manifest itself in lower instantaneous SFRs during individual star formation episodes \citep[e.g.][]{Ward2022}. Rather, the indicators of quenching correlate more strongly with the BH mass, which is a measure of the cumulative output of the AGN over the lifetime of the galaxy \citep[e.g.][]{Piotrowska2022}. While measuring BH masses is beyond the scope of this study, we note that our results do not argue against AGN feedback having an influence on star formation in the dwarf regime. Indeed, recent work \citep[e.g.][]{Davis2022} has suggested that the central gas reservoirs in dwarfs are much larger than what is required to fuel the black hole and the energetics suggest that AGN feedback is both plausible and likely in these systems.

\section{Summary}
\label{sec:summary}

We have employed SED fitting via \textsc{Prospector} using deep UV to MIR broadband photometry in the COSMOS field, to search for radiatively-efficient AGN in nearby ($0.1<z<0.3$) dwarf (10$^{8}$ M$_{\odot}$ < $M_{\rm{\star}}$ < 10$^{10}$ M$_{\odot}$) galaxies. We have explored the percentage of nearby dwarfs that show signs of AGN activity, investigated the potential role of environment and interactions in triggering our AGN and studied whether the sSFRs of our AGN are different from their non-AGN counterparts. Our main conclusions are as follows:

\begin{itemize}
    
    \item Around a third (32 $\pm$ 2 per cent) of dwarf galaxies in our sample show signs of AGN activity, consistent with the dwarf AGN fractions derived, via various techniques, in the recent literature.  

    \item The local environments of dwarf AGN, parameterised by the projected distances to nodes, filaments and massive galaxies, are indistinguishable from those of a control sample (which has the same distribution of redshift and stellar mass) constructed from non-AGN dwarfs. This suggests that local environment is not likely to play a significant role in AGN triggering. 
        
    \item The early-type and interacting fractions in our AGN are also indistinguishable from those of a control sample, suggesting that the onset of AGN, at least in the sample studied here, is not strongly influenced by galaxy interactions. 
        
    \item Given the median values of sSFR and their uncertainties, the star formation activity in dwarf AGN is marginally lower compared to those in their control counterparts. However, we do not find strong evidence of a significant, prompt quenching of star formation in dwarf galaxies which host AGN activity. 
        
\end{itemize}

Our study adds to the burgeoning body of work which suggests that a significant minority of nearby dwarf galaxies show signs of AGN activity. It also demonstrates the potential of SED fitting for exploring AGN populations within dwarfs in future multi-wavelength deep-wide datasets. These studies will be important, both for putting such results on a firmer statistical footing, and for exploring how the AGN populations in dwarfs may evolve over cosmic time. 

%The AGN fraction provides a lower limit to the occupation fraction of black holes in dwarf galaxies and is a useful constraint on predictions of black hole occupancy and AGN fractions in cosmological simulations like NewHorizon.  

%Future endeavours will focus on exploring AGN fractions in different morphological classes within the dwarf galaxy population, studying whether AGN fractions are enhanced in interacting dwarfs and utilizing the AGN populations produced by this study to train machine learning algorithms to find AGN in large surveys like LSST. 

%.............................................................

\section*{Acknowledgements}
We thank the referee for several constructive comments that improved the presentation of the paper. BB and IL acknowledge PhD studentships from the Centre for Astrophysics Research at the University of Hertfordshire. SK and AEW acknowledge support from the STFC (grant numbers ST/S00615X/1 and ST/X001318/1). SK also acknowledges a Senior Research Fellowship from Worcester College Oxford. SD acknowledges support from the STFC via grant ST/W507490/1. DJBS acknowledges support from the STFC via grant ST/V000624/1. For the purpose of open access, the author has applied a Creative Commons Attribution (CC BY) licence to any Author Accepted Manuscript version arising from this submission. 

%.............................................................

\section*{Data Availability}
The parameters calculated in this study can be shared upon reasonable request to the authors. 

%.............................................................

\bibliographystyle{mnras}
\bibliography{dagn}

%.............................................................

% \appendix

% \section{Density Maps}
% \label{sec:density_maps_appendix}

% \begin{figure*}
% 	\includegraphics[width=\textwidth]{images/results/CLASSICDensityMap_1_4_1.pdf}
%     \caption{\textsc{DisPerSE} generated density map for redshift limit $0.18 < z < 0.28$ with dwarfs that host AGN (red) and a control non-AGN sample matched in stellar mass (black) shown over-plotted. The filaments are shown using solid black lines. {\color{green}Why are there so few galaxies in the last density map? I would expect this map to have the largest number of galaxies. The redshift limits dont look right. The first density map is not needed because it is already in the text. All panels should have the same size.}}
%     \label{fig:density_maps1}
% \end{figure*}

% \begin{figure*}
% 	\includegraphics[width=\textwidth]{images/results/CLASSICDensityMap_5_8_1.pdf}
%     \caption{\textsc{DisPerSE} generated density map for redshift limit $0.27 < z < 0.31$ with dwarfs that host AGN (red) and a control non-AGN sample matched in stellar mass (black) shown over-plotted. The filaments are shown using solid black lines.}
%     \label{fig:density_maps2}
% \end{figure*}

%.............................................................

% Don't change these lines
\bsp	% typesetting comment
\label{lastpage}
\end{document}